\documentclass[letterpaper,twocolumn,showpacs,
prl 
,superscriptaddress
]{revtex4}

\usepackage{graphicx}
\usepackage{amssymb}
\usepackage{amsmath,amsfonts,latexsym}
\usepackage{array,tabularx,color}
\usepackage{dcolumn}               
\usepackage[normalem]{ulem}
\usepackage{comment}
\usepackage[amssymb]{SIunits}

\newcommand{\vect}[1]{{\mathbf #1}}

\begin{document}


\title{Propagating wave-packets and quantised currents in coherently
  driven polariton superfluids}

\author{M. H. Szyma\'nska}
\affiliation{Department of Physics, University of Warwick, Coventry,
  CV4 7AL, UK}
\altaffiliation{also at London Centre for Nanotechnology, UK}

\author{F. M. Marchetti}
\affiliation{Departamento de F\'isica Te\'orica de la Materia
Condensada, Universidad Aut\'onoma de Madrid, Madrid 28049, Spain}

\author{D. Sanvitto} 
\affiliation{Departamento de F\'isica de
Materiales, Universidad Aut\'onoma de Madrid, Madrid 28049, Spain}

\date{May 25, 2010}

\begin{abstract}
  We study the properties of propagating polariton wave-packets and
  their connection to the stability of doubly charged
  vortices. Wave-packet propagation and related photoluminescence
  spectra exhibit a rich behaviour dependent on the excitation regime.
  We show that, because of the non-quadratic polariton dispersion,
  doubly charged vortices are stable only when initiated in
  wave-packets propagating at small velocities. Vortices propagating
  at larger velocities, or those imprinted directly into the polariton
  optical parametric oscillator (OPO) signal and idler are always
  unstable to splitting.
\end{abstract}




 
 







\maketitle

After almost a decade since the first reports of bosonic stimulation
\cite{savvidis00:prl,stevenson00}, resonantly pumped polaritons have
been shown to exhibit a new form of non-equilibrium superfluid
behaviour~\cite{amo09,amo09_b,sanvitto09}. The experimental ability to
control externally the polariton currents (phase variations) with the
laser pump~\cite{amo09_b} as well as to generate propagating polariton
wave-packets (density and phase variations), by applying another
pulsed laser probe~\cite{amo09}, is a unique feature of 
driven polariton systems. Here, the interplay between phase and
amplitude variations opens the possibility for novel superfluid
phenomena.
In addition, by using a pulsed Laguerre-Gauss beam, polaritons in the
optical parametric oscillator (OPO) regime have been recently shown to
display persistence of currents and metastability of quantum vortices
with charge $m=1$ and $m=2$~\cite{sanvitto09}.

In this Letter we report a comprehensive theoretical analysis of
propagating polariton wave-packets, triggered by a pulsed probe, and
examine the stability of vortices of charge $m=2$ in different
excitation regimes. In particular, we asses how the wave-packet shape,
velocity and intensity, as well as the stability of $m=2$ vortices,
are related to the probe wavevector and to the photoluminescence (PL)
spectra.
The behaviour of $m=2$ vortices has been studied
recently~\cite{sanvitto09}, and two distinct regimes have been singled
out: In the TOPO regime the initiated vortex only lasts for as long as
the additional decaying gain introduced by the triggering probe. Here,
$m=2$ vortices have been shown to be remarkably stable when initiated
at small group velocities and instead to split into two $m=1$ vortices
at larger velocities. In a second regime, vortices have been shown to
withstand the gain and imprint into the OPO. In this case, $m=2$
vortices always split. We provide the theoretical explanation for this
intriguing phenomena.

The behaviour of $m=2$ vortices has been the subject of intensive
research in the context of ultra-cold atomic gases.  However, although
stable free $m=2$ vortices have been predicted for specific ranges of
density and interaction strength~\cite{pu99,machida03}, they have not
been observed experimentally~\cite{shin04}. Stable pinned $m=2$
persistent currents have been recently
realised~\cite{phillips07} only by using toroidal traps.

\paragraph{Model}
The time evolution of coherently driven polaritons is described by the
Gross-Pitaevskii equations for coupled cavity photon and exciton
fields $\psi_{C,X} (\vect{r},t)$ with pump and decay
($\hbar=1$)~\cite{whittaker2005_b}:
\begin{multline}
  i\partial_t \begin{pmatrix} \psi_X \\ \psi_C 
\end{pmatrix}
  = \begin{pmatrix} 0 \\ F_p + F_{pb} \end{pmatrix}
  \\ + \begin{pmatrix} \omega_X -i \kappa_X + g_X|\psi_X|^2& \Omega_R/2
    \\ \Omega_R/2 & \omega_C -i \kappa_C \end{pmatrix} \begin{pmatrix}
    \psi_X \\ \psi_C
  \end{pmatrix}\;.
\label{eq:model}
\end{multline}
We use here the same notation as in Ref.~\cite{marchetti10}. The
cavity field is driven by a continuous-wave pump, $F_p(\vect{r},t) =
\mathcal{F}_{f_p,\sigma_p} (r) e^{i (\vect{k}_p \cdot \vect{r} -
  \omega_p t)}$
with a Gaussian or a top-hat spatial profile
$\mathcal{F}_{f_p,\sigma_p}$ of strength $f_p$ and full width at half
maximum (FWHM) $\sigma_p$. $F_{pb}(\vect{r},t)$ is the pulsed probe
introduced later. We neglect the exciton dispersion and assume a
quadratic dispersion for photons, $\omega_C=\omega_C^{0}
-\frac{\nabla^2}{2m_C}$.  $\Omega_R$ is the Rabi splitting and the
fields decay with rates $\kappa_{X,C}$.  The exciton interaction
strength $g_X$ is set to one by rescaling both fields $\psi_{X,C}$ and
pump strength $F_{p}$ and $F_{pb}$ by $\sqrt{\Omega_R/(2
  g_X)}$. Through the paper, $m_C=2 \times 10^{-5} m_0$, the energy
zero is set to $\omega_X=\omega_C^{0}$ (zero detuning) and $\Omega_R =
4.4$meV. We numerically solve Eq.~\eqref{eq:model} on a 2D grid using
5$^{\text{th}}$-order adaptive-step Runge-Kutta algorithm.

\begin{figure*}
\centering
\includegraphics[width=\textwidth]{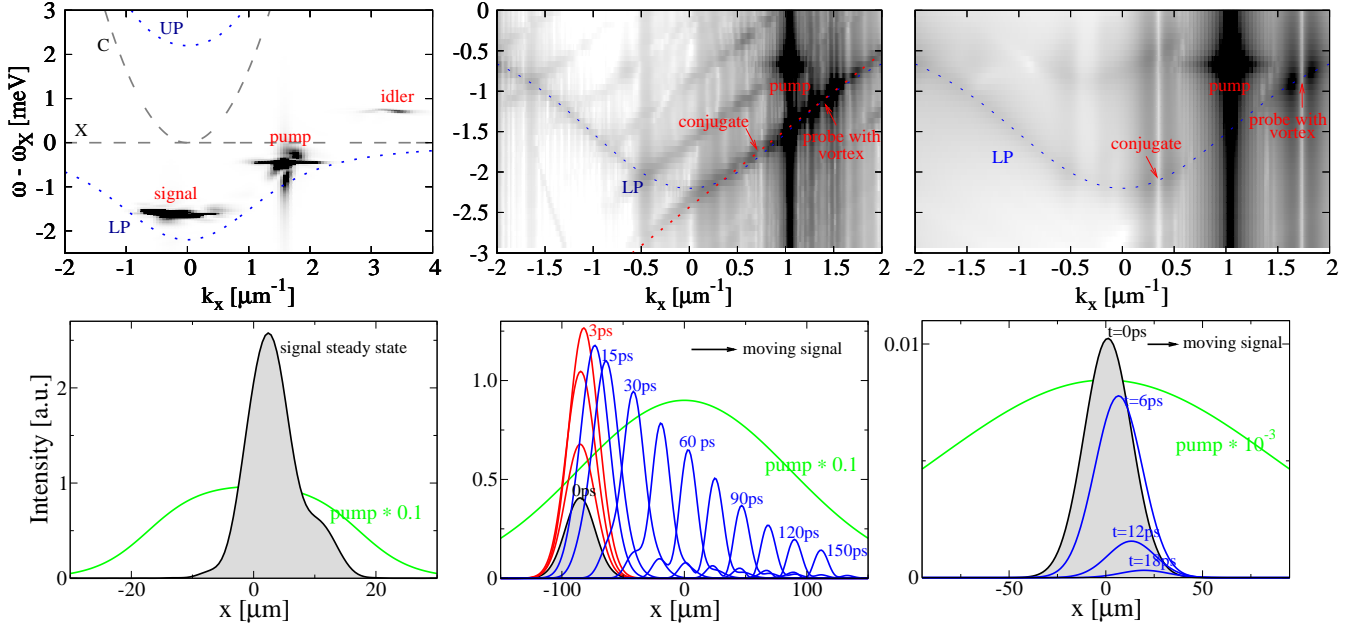}
\caption{(Color online) PL spectra (top) and spatial profiles of pump
  and filtered signal $|\psi_C^{s}(x,0,t)|$ (bottom) for three
  regimes: (i) OPO (left panels) for a smoothen top-hat pump above
  threshold ($f_p=1.12 f_p^{\text{(th)}}$): polaritons, continuously
  injected at $(k_p,0)$ and $\omega_p$, undergo stimulated scattering
  into signal and idler stationary steady states
  ($\kappa_{X,C}=0.2$meV). (ii) TOPO (middle panels): a short,
  $\sigma_t=1$ps, Laguerre-Gauss $m=2$ (top) and Gaussian $m=0$
  (bottom) probe~\eqref{eq:probe} resonant with $(k_{pb},0)$ and
  $\omega_{pb}$ triggers propagating signal and conjugate states,
  which lock to the same group velocity ($\kappa_{X}=0$,
  $\kappa_{C}=0.02$meV). (iii) Similar conditions to (ii) but
  $\kappa_{X,C}=0.09$meV (right panels); now there is no amplification
  of the probe, both signal and conjugate decay exponentially. Lower
  (LP) and upper (UP) polariton (blue dashed lines), and bare cavity
  photon (C) and exciton (X) dispersions (gray dashed lines) are shown
  for reference.}
\label{fig:spect}
\end{figure*}
%
%
In the absence of the probe ($F_{pb}=0$), polaritons are continuously
injected into the \emph{pump} state and, above the pump strength
threshold for OPO, $f_p^{\text{th}}$, undergo stimulated scattering
into the \emph{signal} and \emph{idler} states~\cite{stevenson00} (see
Fig.~\ref{fig:spect} left panels). We filter the signal, idler and
pump spatial profiles
$|\psi_{C,X}^{s,i,p}(\vect{r},t)|e^{i\phi_{C,X}^{s,i,p}(\vect{r},t)}$
in a cone around their associated momenta $\vect{k}_{s,i,p}$. The
supercurrent, $\nabla \phi_{C,X}^{s,i,p}$, is a superposition of the
dominant uniform flow $\vect{k}_{s,i,p}$ and more complex currents
(caused by the system being finite size), which move particles from
  gain to loss dominated regions.
In addition one can characterise the density variations defining the
\emph{group velocity} $\vect{v}^{s,i,p}_g= d \vect{r_m^{s,i,p}} /dt$,
where $\vect{r_m^{s,i,p}}$ is the maximum of the signal (idler and
pump) spatial profile. The OPO is a steady state regime, where
$|\psi_{C,X}^{s,i,p}(\vect{r},t)|$ are time independent. This is also
recognisible in the typical flat dispersion of signal, idler and pump
PL spectra (see left panel of Fig.~\ref{fig:spect}) --- the signal
(idler, pump) group velocity being the derivative of the dispersion at
$\vect{k}_{s}$ ($\vect{k}_{i,p}$).

\paragraph{Propagating wave-packets and TOPO regime}
In order to create propagating wave-packets (finite group velocity),
one needs to add a pulsed probe with finite momentum $\vect{k}_{pb}$,
spatially small with respect to the pump spot size~\cite{amo09}. Since
we are interested in the stability of triggered vortices of charge
$m=2$, and its connection to the character of the propagating
wave-packet, we consider here a Laguerre-Gauss pulsed
probe~\cite{whittaker07,sanvitto09,marchetti10},
\begin{multline}
  F_{pb}(\vect{r},t) = f_{pb} |\vect{r}-\vect{r}_{pb}|^{|m|}
  e^{-|\vect{r}-\vect{r}_{pb}|^2/(2\sigma^2_{pb})} e^{i m \varphi} \\
  \times e^{i (\vect{k}_{pb} \cdot \vect{r} - \omega_{pb} t)}
  e^{-(t-t_{pb})^2/(2\sigma^2_{t})} \; ,
\label{eq:probe}
\end{multline}
where the phase $\varphi$ winds from $0$ to $2\pi$ around the vortex
core $\vect{r}_{pb}$. Above threshold for OPO, $f_p> f_p^{\text{th}}$,
one can identify two distinct regimes depending on the system
parameters~\cite{sanvitto09}: In one case, the triggered vortex
propagates as extra population (gain) on top of the OPO stationary
signal and idler states~\cite{amo09} --- TOPO regime.  Here, the
vortex lasts only for as long as the additional decaying gain. A
similar TOPO regime can be found also below the OPO threshold, and for
simplicity we show this case in Fig.~\ref{fig:spect} (middle panel).
Alternatively, vortices persist the gain and get imprinted into the
OPO signal and idler --- metastable vortex regime~\cite{marchetti10}.

%
In the TOPO regime, the polaritons generated by the probe scatter
parametrically with the ones of the pump, generating a traveling
signal (here meant as extra population on top of either the OPO signal
or idler, or full signal if below the OPO threshold) with momentum
$\vect{k}_s=\vect{k}_{pb}$ (which can be either $\vect{k}_{pb} >
\vect{k}_{p}$ or $\vect{k}_{pb} < \vect{k}_{p}$) and its
\emph{conjugate} at $\vect{k}_c=2\vect{k}_p-\vect{k}_{pb}$.
In the case of the middle panels of Fig.~\ref{fig:spect} the signal
and the conjugate get initially strongly amplified by the parametric
scattering from the pump state, then decay slowly (we refer to this as
a proper TOPO regime) and, at later times, eventually the decay
becomes exponential. This regime reproduces the one observed in
experiments (see Fig. 3 of Ref.~\cite{tosi10}), where the intensity
maximum is reached quickly, within $4$ps after the maximum of the
pulsed probe, and is immediately followed by a slow decay. Instead,
for the case of the right panels of Fig.~\ref{fig:spect}, parametric
scattering is too weak to induce any significant amplification, and we
observe exponential decay of signal and conjugate populations
immediately after the probe switches off.

\begin{figure}
\begin{center}
\includegraphics[width=1.0\linewidth,angle=0]{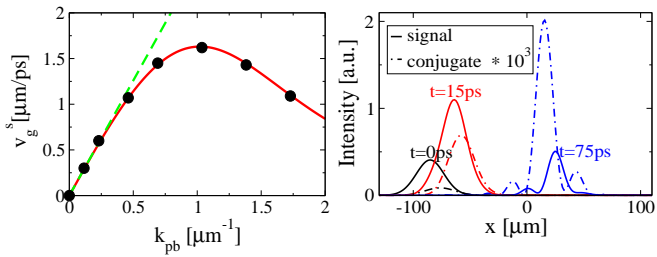}
\end{center}
\caption{(Color online) Left panel: Group velocity, $v^s_g$ of the
  propagating signal wave-packet as a function of the probe momentum
  $k_{pb}$. The black dots are determined from simulations, whereas
  the solid line (red) is the derivative of the LP dispersion
  evaluated at $k_{pb}$, $v^{LP}_{k_{pb}}$. The green dashed line
  indicates where the LP dispersion deviates from the quadratic
  behaviour. Right panel: Signal (solid lines) and conjugate (dashed
  lines) states for $k_{pb}= 1.4\mu$m$^{-1}$ at different times after
  the arrival of the probe.}
\label{fig:group}
\end{figure}

By analysing the change in time of the spatial profile of the TOPO
signal $|\psi_{C,X}^{s}(\vect{r},t)|$, we find that its group velocity
$v_g^{s}$ is simply given by the derivative of the lower polariton
(LP) dispersion evaluated at $\vect{k}_{pb}$, i.e. for zero detuning
and low densities by $v^{LP}_{k_{pb}} \equiv k_{pb}/(2m_C)-k^3/(2m_C
\sqrt{k^4 + 4m_C^2 \Omega_R^2})$ (see Fig.~\ref{fig:group}). We have
confirmed this for both TOPO regimes.

In order to understand this result it is instructive to analyse the PL
spectra. In the regime with weak parametric scattering (right panel of
Fig.~\ref{fig:spect}), aside the strong emission from the pump state,
the dispersion is simply that of the LP, and thus the signal
propagates with a group velocity given by $v^{LP}_{k_{pb}}$. However,
in the proper TOPO regime (middle panel of Fig.~\ref{fig:spect}) the
spectrum becomes linear.
This can be explained by the fact that in order to have efficient
parametric scattering processes, signal and conjugate require a large
spatial overlap and so their group velocities need to lock to the same
value, therefore the dispersion becomes linear --- a similar result
has been found in 1D simulations, as well as in
experiments~\cite{amo09}. Here, we determine that the TOPO linear
dispersion is tangential to the LP branch at $k_{pb}$, thus its slope
is given in this case also by $v^{LP}_{k_{pb}}$.

The linearisation of the TOPO dispersion, with the slope defined by
the probe momentum $k_{pb}$, also explains our observation that the
TOPO amplification is stronger when signal
($\vect{k}_{s}=\vect{k}_{pb}$) and conjugate ($\vect{k}_{c}$)
wavevectors (connected via parametric constraint
$2\vect{k}_p=\vect{k}_{s}+\vect{k}_c$) have nearby values.
This implies that the LP tangential line at $\vect{k}_{pb}$ lies close
to the LP dispersion also at $\vect{k}_{c}$, therefore leading to
efficient stimulation.
It is interesting to note, that by changing $\vect{k}_p$ and
$\vect{k}_{pb}$ it is possible to engineer wave-packets with a broad
range of values of group velocities and supercurrents --- e.g., for
$\vect{k}_{pb}=2\vect{k}_p$, one has a conjugate state at
$\vect{k}_c=0$ (i.e., no net current) with a large group velocity,
$v^c_g=v^{LP}_{2k_p}$.

From the PL spectrum we can deduce the nature of the wave-packet
propagation. For linear dispersion one expects a soliton-like
behaviour, with signal and conjugate propagating and not changing
neither shape nor intensity.
For quadratic dispersion, propagation is analogous to the ballistic
time-of-flight expansion of ultra-cold atomic
gases~\cite{pitaevskii03}, i.e. for a Gaussian wave-packet, the
FWHM=$(\sigma^2_{pb}+(\frac{t}{2m_C \sigma_{pb}})^2)^{1/2}$ grows in
time, but the shape is preserved.
The total density decays exponentially with rate given by
$\frac{\kappa_C+\kappa_X}{2}$ at zero detuning, and the maximum of the
wave-packet moves with a constant velocity $v^{LP}_{k_{pb}}$ (equal to
$\frac{\vect{k}_{pb}}{2m_C}$ in the quadratic LP regime).
Due to the dynamical nature of the TOPO state, the system evolves
between different regimes. In particular, only in the strong
amplification regime the spectrum is linear, while it evolves back to
LP at longer times. In addition, when the LP dispersion deviates from
quadratic, propagation becomes complex: the wave-packet gets distorted
and one observes beatings in the spatial profiles.

Propagation of signal and conjugate are shown in the right panel of
Fig.~\ref{fig:group}. We observe a mechanism analogous to the one
found in four-wave-mixing experiments~\cite{boyer07,boyer08}: when the
probe arrives the conjugate propagates faster then the signal, before
getting locked to it with a small spatial shift of their maximum
intensities. At later times, when the density drops and the parametric
process becomes inefficient, the two wave-packets start unlocking ---
the conjugate slows down with respect to the signal if $k_c<k_s$ as in
Fig.~\ref{fig:group}, or moves faster when $k_c>k_s$.

\begin{figure}
\begin{center}
\includegraphics[width=1.0\linewidth,angle=0]{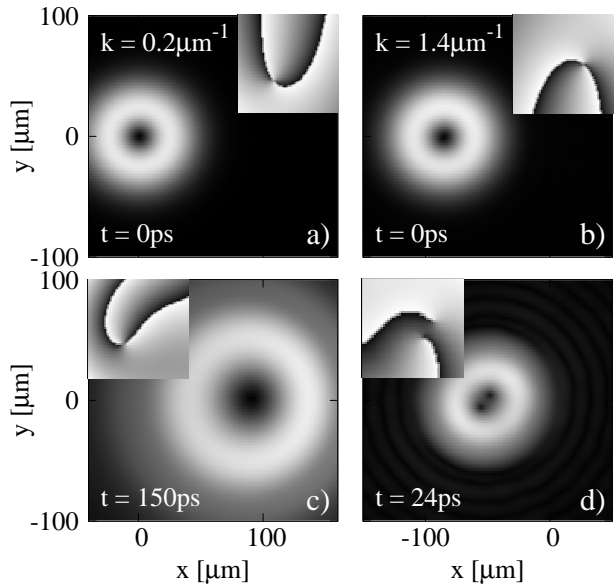}
\end{center}
\caption{Profiles of TOPO signal intensity $|\psi_C^s (\vect{r},t)|$
  and phase $\phi_{C}^{s}(\vect{r},t)$ (inset) after the arrival (at
  $t=0$) of an $m=2$ vortex pulsed probe~\eqref{eq:probe} with
  $\sigma_{pb} \simeq 87\mu$m. The conditions are the same as the
  middle panels of Fig.~\ref{fig:spect}, but with
  $k_{pb}=0.2\mu$m$^{-1}$ (a,c) and $k_{pb}=1.4\mu$m$^{-1}$
  (b,d). While in the case (a,c) the $m=2$ vortex does not split
  within its lifetime, in (b,d) the vortex splits soon after the probe
  arrives.  The intensity scale in (c) is 200 times smaller then in
  (a) -- signal expands as the density drops two orders of magnitude
  (see text).}
\label{fig:vortex}
\end{figure}
%
\paragraph{Stability of $m=2$ TOPO vortices}
In the TOPO regime, we observe $m=2$ vortices to be stable within
their lifetime when triggered at small momenta $k_{pb}$
(Fig.~\ref{fig:vortex}a,c), while TOPO vortices always split into two
$m=1$ vortices for large $k_{pb}$ (Fig.~\ref{fig:vortex}b,d), in
agreement with the recent experimental
observations~\cite{sanvitto09}. Our numerical analysis shows that the
crossover from non-splitting to splitting happens for values of the
probe momentum where the LP dispersion deviates from quadratic
(see Fig.~\ref{fig:group}).
The two regimes are shown in Fig.~\ref{fig:vortex}: For
$k_{pb}$=0.2$\mu$m$^{-1}$, at short times, the signal carrying the
vortex propagates without changing shape and with little change in
intensity, consistent with the corresponding PL spectrum having a
linear dispersion. However, at longer times the density drops more
then two orders of magnitude, the dispersion becomes quadratic and
therefore the wave-packet expands (panel c). A uniform expansion of
the wave-packet leads to the increase of the vortex core but does not
cause the vortex to split. In contrast, for $k_{pb}$=1.4$\mu$m$^{-1}$,
where the LP dispersion is not quadratic, the splitting of the $m=2$
vortex state into two $m=1$ vortices happens shortly after the arrival
of the probe (panel d).
This behaviour can be explained as follows: the dispersion of the
time-dependent TOPO evolves from LP (before the probe arrival) to
linear (at early times after the probe arrival), and back to the LP
dispersion at later times.  Wave-packets propagating with
non-quadratic dispersion do not keep their shapes. Indeed, we see the
distortion to be very pronounced at later times of the evolution. This
distortion during the propagation leads to the mechanical splitting of
$m=2$ vortex, analogous to the structural instability discussed
in~\cite{garciaripoll01}. Also, as discussed in \cite{sanvitto09}, for
small $k_{pb}$, within the quadratic part of the dispersion, the group
velocity of the propagating vortex and the velocity associated with
the net supercurrent (given by $k_{pb}$) are equal. However, this is
not the case for larger $k_{pb}$, beyond the quadratic part, and so
the propagating vortex feels rather strong net current in its frame,
which may provide additional explanation for splitting.

\begin{figure}
\begin{center}
\includegraphics[width=1.0\linewidth,angle=0]{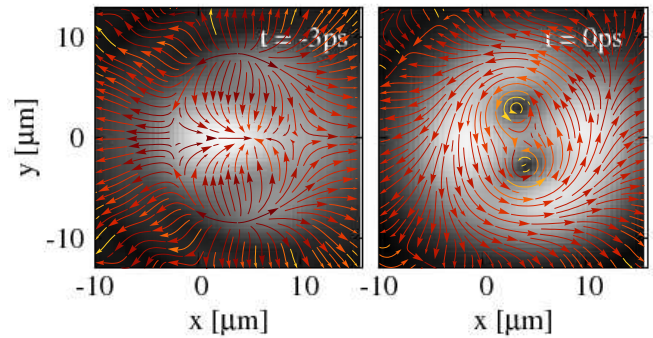}
\end{center}
\caption{ (Color online) Filtered signal profile and currents above
  OPO threshold $f_p=1.12 f_p^{\text{(th)}}$ for a top-hat pump with
  FWHM $\sigma_p=35\mu$m at $t=-3ps$ (left panel) before the arrival
  of an $m=2$ probe at $t=0$ps (right). The double quantised vortex
  splits into two $m=1$ vortices even before the probe reaches it's
  maximum intensity.}
\label{fig:vortexOPO}
\end{figure}
%
\paragraph{$m=2$ vortices imprinted into the OPO}
As in Ref.~\cite{sanvitto09}, we find that doubly charged ($m=2$)
vortices which get imprinted into the OPO signal are never stable, and
split into two $m=1$ vortices almost immediately, even before the
probe reaches its maximum intensity, as shown in
Fig.~\ref{fig:vortexOPO}. There are several causes of the splitting:
Before the probe arrives, the OPO dispersion is flat around the
signal, idler, and pump. However, the triggering probe favours the
signal and conjugate to lock and propagate with the same velocity
$v^{LP}_{k_{pb}}$, which corresponds to a linear dispersion. Further,
once the vortex gets imprinted into the stationary OPO signal and
idler, the dispersion changes back to be flat.  The evolution of the
dispersion between linear and flat leads to a complicated dynamics of
both signal and idler (the \emph{transient period} described
in~\cite{marchetti10}), causing the structural instability and
splitting of the vortex during the transient time. Another reason for
the structural instability during the vortex imprinting into the OPO
signal and idler are the non-uniform currents (shown in
Fig.~\ref{fig:vortexOPO}) caused by the interplay between spatial
inhomogeneity, pump and decay, which the OPO vortex experiences in its
reference frame.

To conclude, we have determined group velocity and supercurrents of
propagating polariton wave-packets triggered by a short pulsed laser
probe, and established the conditions under which triggered $m=2$
vortices are stable and when they split into two vortices of
$m=1$. Both phenomena are related to the form of the PL spectrum as
well as the wave-vector of the triggering probe.
The ability to control the topological charge and number of vortices,
simply by changing the wave-vector of the external probe, holds
potential for operations in quantum information. In addition, we have
shown that, by varying the pump and probe wavevectors, it is possible
to engineer polariton wave-packets with any relation between group and
supercurrents velocities --- a property which can be useful for
storage applications.

\acknowledgments We are grateful to E. Cancellieri, J. Keeling,
C. Tejedor and L. Vi\~na for stimulating discussions. F.M.M. and D.S.
acknowledge financial support from the programs Ram\'on y Cajal and
INTELBIOMAT (ESF).  We thank TCM group (Cavendish Laboratory,
Cambridge) for the use of computer resources.


\newcommand\textdot{\.}

\end{document}